\documentclass[12pt,a4paper]{article}
\usepackage[utf8x]{inputenc}
\usepackage{a4wide}
\usepackage{amsmath,amssymb,amstext,amsthm,amssymb,amsfonts}
\usepackage[nosort]{cite}
\usepackage{hyperref}
\usepackage{graphicx}
\usepackage{color}
\usepackage{slashed}

\newcommand{\Pl}{\mathrm{P}}
\newcommand{\diff}{\mathrm{d}}
\newcommand{\tr}{\mathrm{tr\,}}

\title{Towards a Swampland Global Symmetry Conjecture using Weak Gravity}
\author{Tristan Daus, Arthur Hebecker, Sascha Leonhardt, John March-Russell}

\begin{document}
 
 \thispagestyle{empty}
 \vspace*{0.1cm}
 \begin{center}
  {\Large\bf 
   Towards a\\\vspace*{.2cm}
   Swampland Global Symmetry Conjecture\\\vspace*{.4cm}
   using Weak Gravity
  }
 
  \vspace*{1.0cm}
 
  {\large
   Tristan Daus$,\!^{1}$ Arthur Hebecker$,\!^{1}$ Sascha Leonhardt\,$\!^{1}$\\[2mm] and John March-Russell\,$\!^{2}$
  }\\[0.6cm]

  {\it
   $^1\,$Institute for Theoretical Physics, Heidelberg University,\\ Philosophenweg 19, 69120 Heidelberg, Germany\\[3mm]
   $^2\,$Rudolf  Peierls  Centre  for  Theoretical  Physics,  University  of  Oxford,\\  Beecroft Building,  Oxford  OX1 3PU, United Kingdom\\[3mm]
   {\small\tt (\,t.daus, a.hebecker, s.leonhardt~@thphys.uni-heidelberg.de\,,\\ John.March-Russell~@physics.ox.ac.uk \,)}
  }\\[1.5cm]
 \end{center}
 
 \begin{abstract}\normalsize
  It is widely believed and in part established that exact global symmetries are inconsistent with quantum gravity. One then expects that approximate global symmetries can be \emph{quantitatively} constrained by quantum gravity or swampland arguments. We provide such a bound for an important class of global symmetries: Those arising from a gauged $U(1)$ with the vector made massive via Higgsing with an axion. The latter necessarily couples to instantons, and their action can be constrained, using both the electric and magnetic version of the axionic weak gravity conjecture, in terms of the cutoff of the theory.  As a result, instanton-induced symmetry breaking operators with a suppression factor not smaller than $\exp(-M_\Pl^2/\Lambda^2)$ are present, where $\Lambda$ is a cutoff of the 4d effective theory. We provide a general argument and clarify the meaning of $\Lambda$. Simple 4d and 5d models are presented to illustrate this, and we recall that this is the standard way in which things work out in string compactifications with brane instantons. The relation of our constraint to bounds that can be derived from wormholes or gravitational instantons and to those motivated by black-hole effects at finite temperature are discussed, and we present a generalization of the Giddings-Strominger wormhole solution to the case of a gauge-derived $U(1)$ global symmetry. Finally, we discuss potential loopholes to our arguments.
 \end{abstract}
 \newpage

 \section{Introduction}
  
  It is common lore that a quantum field theory, if consistently embedded in quantum gravity, will not possess exact global symmetries \cite{Banks:1988yz, Giddings:1987cg, Lee:1988ge, Abbott:1989jw, Coleman:1989zu, Kamionkowski:1992mf, Holman:1992us, Kallosh:1995hi, Banks:2010zn, Harlow:2018tng}. The standard argument invokes black hole evaporation, in which the global charge hidden behind the horizon simply disappears (see however \cite{Dvali:2016mur,MarchRussell:2002fn} for a discussion of subtleties related to topological charges). But it is not straightforward to translate this to a quantitative statement about symmetry-breaking operators in the low-energy effective theory.  We attempt to address this question in an important class of models: Those possessing a linearly realized, approximate global symmetry which derives from a $U(1)$ gauge theory. 

  Of course, the size of coefficients of global-symmetry-violating operators has been discussed for a long time on the basis of wormholes or, more generally, gravitational instantons \cite{Lee:1988ge, Abbott:1989jw, Coleman:1989zu}.  Moreover, in the case of a spontaneously broken global $U(1)$, an axion exists.  Symmetry breaking is then encoded in the instanton-induced axion potential, which is constrained using the axionic version of the Weak Gravity Conjecture (WGC) \cite{ArkaniHamed:2006dz}  (for recent work in this direction see e.g.~\cite{Cheung:2014vva, delaFuente:2014aca, Rudelius:2015xta, Montero:2015ofa, Brown:2015iha, Bachlechner:2015qja, Hebecker:2015rya, Junghans:2015hba, Heidenreich:2015wga, Kooner:2015rza, Kaloper:2015jcz, Kappl:2015esy, Choi:2015aem, Saraswat:2016eaz, Palti:2019pca}).  By contrast, our focus here is on linearly realized global symmetries, which can e.g.~be used to protect some type of particle number in the low-energy effective field theory (EFT).  In specific cases, relevant constraints deriving from the WGC have recently been given in \cite{Hebecker:2019vyf, Fichet:2019ugl}.  Additionally, a general bound, independent of the WGC but rather motivated by black hole effects in a thermal plasma, has been conjectured in \cite{Fichet:2019ugl}.  Since it is likely that gauge symmetries are also constrained by swampland arguments, e.g.~the total rank of the gauge group, and our interest here is in global symmetries, we here adopt the terminology Swampland Global Symmetry Conjecture for our statements and bounds, but we will argue that the precise formulation and underpinning of the conjecture is yet to be determined.  We will make a corresponding suggestion.

  Our main technical result goes beyond previous work as follows: First, we claim that given a slight, natural generalization of the WGC and the completeness hypothesis our constraint can actually be derived. Second, while not completely general, it addresses a very large class of constructions which play a central role in model building in general and in particular in string compactifications \cite{Blumenhagen:2009qh}. The models we want to consider have an underlying gauged $U(1)$ symmetry.  If this $U(1)$ is non-linearly realized, the vector and the Nambu-Goldstone boson or axion are removed from the spectrum.\footnote{
   Here we are interested in the case where the vector mass is not parametrically below the cutoff of the theory, so we are not considering the limit of small $U(1)$ gauge coupling, and the connection with the physics of light vector states with St\"uckelberg masses~\cite{Reece:2018zvv,Craig:2018yld}, or anomalies~\cite{Craig:2019zkf}.   
  }
  The axion may be a fundamental periodic scalar or the phase of a complex Higgs, though there may be some differences as we explain below. Importantly if some of the originally $U(1)$-charged particles survive in the low-energy effective theory, they will transform under a global $U(1)$.  The latter is linearly realized, in spite of the fact that the high-scale gauge $U(1)$ is removed.  The reason is simply that the axion is not part of the low-energy theory. 

  As we argue in the following, crucially, the axion should couple to some form of instanton -- this is required by the completeness hypothesis \cite{Polchinski:2003bq,Banks:2010zn}. A mild generalization of the WGC to this case further constrains their action \cite{ArkaniHamed:2006dz} and, in its magnetic form for axions \cite{Hebecker:2017uix},\footnote{
   This bound has also been used to constrain a St\"uckelberg mass~\cite{Reece:2018zvv}, which is however not our interest in the present paper.
  }
  provides a relation to the UV cutoff of the 4d effective theory. Moreover, as will be argued in full generality below, these instantons necessarily induce EFT operators which violate the global $U(1)$.  This leads to the desired quantitative bound.

  As an interesting fact we note that, while very different in their motivation and range of applicability, all of the above bounds on symmetry-violating operator coefficients have the parametric form $\exp(-M_\Pl^2/\Lambda^2)$. In all cases, from wormholes to instantons to black holes in a thermal plasma, one may argue that the parametrics are necessarily the same: The exponent is simply the Einstein-Hilbert action of some localized object, with $\int \diff^4x\,R$ replaced by $1/\Lambda^2$, where $\Lambda$ is the UV cutoff. Specifically in the wormhole context, we find a generalization of the well-known Giddings-Strominger solution to the case of a $U(1)$ gauge-derived global symmetry where a globally charged particle passes through the wormhole.

  Finally, we note, and will discuss in more detail below, that the above parametric similarity suggests a simplicity which might be misleading. First, it is essential whether just one or all symmetry-violating operators must respect the parametric bound above. Second, it may be that different types of approximate global symmetries (to be specified momentarily) call for different bounds. 

  The rest of the paper is organized as follows. In Sect.~\ref{sec:MainArgument}, we review our main idea based on the WGC for axions gauged under a $U(1)$ symmetry and instanton-induced operators, which are symmetry-violating and suppressed by a factor of $\exp(-M_\Pl^2/\Lambda^2)$. Sect.~\ref{sec:SimpleModels} demonstrates this in simple 4d and 5d toy models involving, respectively, fermions and gauge instantons, and a purely bosonic 5d theory compactified to 4d. We discuss some explicit quantum gravity realizations of our bound in Sect.~\ref{sec:QGExamples}, including wormholes and the Euclidean brane instantons of string models. We also comment on recent arguments based on black hole effects in a thermal plasma. Limitations of our approach, in particular a possible loophole related to the numerical coefficient in the exponent, are discussed in Sect.~\ref{lh}. Moreover, assuming that this loophole can be closed or at least its impact limited, we discuss how our results may combine in a general Swampland Global Symmetry Conjecture. We conclude in Sect.~\ref{sec:Conclusion}. Details of a Euclidean wormhole solution containing both 3-form flux and a charged particle are discussed in App.~\ref{sec:AppendixA}.

 \section{Basic argument}\label{sec:MainArgument}
  
 \subsection{Definitions and classification}
  
  Let us start by  defining some basic terminology, without any claim to novelty or originality:  We will say that an EFT possesses an \emph{approximate global symmetry} if among all possible processes, $P(\{i\}\rightarrow \{j\})$, allowed by all spacetime and gauge conservation laws there is a subset $P_\text{gsv}$ with rates that are parametrically smaller, \emph{and that this subset is distinguished by the violation of an otherwise conserved additive or multiplicative quantum number}.\footnote{
   We hope the reader finds the notion ``parametrically smaller rates" intuitively clear!  To precisely define what one means by this is in general involved as can be illustrated by the following example: 
   Consider a 4d $U(1)$ gauge theory with two types of bosons of charge, say, $\pm 1$ and $\pm 11$.  Then the leading gauge-invariant operator that connects the two types of matter in a way that violates the individual (particle - antiparticle) numbers $N_1$ and $N_{11}$ is $\phi_{11} (\phi_1^*)^{11} / \Lambda^8 +\mathrm{h.c.}$  This leads to, e.g., a $\Delta N_1= -11$, $\Delta N_{11}=1$, $2\rightarrow 10$ particle scattering process with cross section parametrically going as $\sigma_{\Delta N}(E) \sim E^{14}/\Lambda^{16}$, where here we are assuming the center-of-mass scattering energy $E\ll \Lambda$ is much greater than the masses of both $\phi_1$ and $\phi_{11}$.  On the other hand there are $\Delta N_1= \Delta N_{11}=0$, $2\rightarrow 10$, particle scattering processes starting with exactly the same initial states which have rates not smaller than $\sigma_{2\rightarrow 10} \sim \alpha^{10}/E^{2}$, where $\alpha$ is the $U(1)$ fine structure constant.  Thus for $E\ll \alpha^{5/8}\Lambda$ the rate of otherwise similar $\Delta N=0$ and $\Delta N\neq 0$ processes is parametrically different. In addition there are many $\Delta N_1= \Delta N_{11}=0$, $2\rightarrow k$ ($k<10$) processes with cross sections $\sigma_{2\rightarrow k}(E) \gg  \sigma_{\Delta N}(E)$, so $\Delta N_{1,11}$ violation is slow.  The issue of almost global symmetries appearing in the EFT due to large ratios of the gauge charges of the light states will be discussed further below.
  }
  Here $\{i\},\{j\}$ label the set of all possible multi-particle initial and final states of the EFT degrees of freedom.  Note that it is important that \emph{all} rates associated with the violation of the relevant additive or multiplicative quantum number are small, for there can be circumstances where individual processes in a theory can be small without there being a good notion of an approximate global symmetry.  (Alternatively, for theories such as conformal field theories which do not have a well-defined notion of particle, we can consider all possible correlation functions of the theory and apply a similar definition.)  In addition we emphasize that the gauge conservation laws may not be associated to long-range massless gauge bosons, as the theory could, and in general will, possess discrete gauge symmetries, either Abelian or non-Abelian which will restrict the allowed processes~\cite{Krauss:1988zc,Alford:1989ch,Preskill:1990bm,Alford:1990pt,Alford:1991vr}.   These discrete gauge symmetries can be distinguished from exact global symmetries by long-range Aharonov-Bohm-type scattering experiments.  
  
  Moreover,  the global symmetries that are of interest to us in this work are associated with, in the continuous case, conventional Noether currents, and, more generally, group action operators faithfully realizing a continuous or discrete group that satisfy certain locality properties.  Such global symmetries are ``splittable" in the terminology of the AdS/CFT proof of Harlow and Ooguri~\cite{Harlow:2018tng}.   Of course, in the approximate global symmetry case the unitary operators enacting the would-be symmetry only approximately commute with the Hamiltonian of the theory and, if we are concerned with a continuous global symmetry, the Noether currents are only approximately conserved.

  Corresponding to this definition the operators in an EFT action describing a theory with an approximate global symmetry may be divided into two disjoint classes: the singlets and the non-singlets with respect to the would-be global group action. Moreover, the non-singlets should either be irrelevant in the Wilsonian sense or have ``small" coefficients (we will later refine the meaning of ``small" and define a notion of a \emph{high-quality} approximate global symmetry).
 
  There are different reasons why an EFT might possess an approximate global symmetry. For example, approximate global symmetries may be 

  \vspace{.2cm}\noindent 
  (1) {\bf Gauge-derived.} With this term we would like to refer to global symmetries following from a non-linearly realized gauge symmetry. Specifically, in the case of a $U(1)$ gauge symmetry Higgsed by an axion both the vector and the pseudoscalar become heavy.  Yet, any charged state which for whatever reason remains light will now be subject to an approximate global $U(1)$ where the coefficients of all symmetry-violating terms in the EFT are small.   We will further explore the physics of such gauge-derived global symmetries as this case will be the main focus of our work.

  \vspace{.2cm}\noindent 
  (2) {\bf Accidental.} Here, spacetime and gauge symmetries, continuous or discrete, forbid all relevant and marginal
  symmetry-violating operators constructed out of the light
  field content of the EFT.\footnote{
   Cf.~$B$ and $L$-symmetry in the Standard Model.  At the level of the relevant and marginal operators there are no terms violating these global symmetries that can be written in the Lagrangian consistent with SM gauge symmetries and Lorentz invariance.  There do exist potential irrelevant operators violating these symmetries. Moreover, given the SM field content there must be irrelevant operators present violating $(B+L)$, but not $(B-L)$, due the 't~Hooft vertex interaction implied by the $U(1)_{B+L}$-$SU(2)_w^2$ anomaly. Note that potential Majorana mass terms for the neutrinos, which would violate $L$ but not $B$, appear to be dimension-three operators in the far IR. However, above the weak scale they can be seen to arise from dimension five operators involving the SM Higgs doublet. This illustrates that the presence of accidental global symmetries depends on the energy scale at which one studies a given model.
  }   
  An interesting WGC-derived bound on how high the mass-dimension of excluded operators can become in simple models has recently appeared in \cite{Fichet:2019ugl} (though it has also been noted that this can be avoided at the price of larger field content). Moreover, in, e.g., the gauged $\mathbb{Z}_N$ case the power of this idea clearly grows with $N$. This $N$, however, may be constrained using black-hole arguments \cite{Dvali:2008tq} or, even more strongly, using also the WGC for 1 and 2-forms~\cite{GarciaGarcia:2018tua, Craig:2018yvw, Fichet:2019ugl}.

  \vspace{.2cm}\noindent
  (3) {\bf Fine-tuned.} By this we mean that the coefficients of all relevant and marginal operators that transform under the would-be global symmetry are ``small" by a landscape-type tuning.  This option is limited in cases where, as expected in string theory, the landscape of EFTs with cutoff $\gtrsim \Lambda$ is finite \cite{Acharya:2006zw,Heckman:2019bzm}. One may try to quantify this by arguing how the number of vacua grows with $M_\Pl/\Lambda$. It is even conceivable that our bound, already advertised in the Abstract and Introduction, is valid for such type-(3) approximate global symmetries for the reason just explained.  In this work, however, we will not be concerned with a quantitative analysis of this interesting possibility.
  \vspace{.2cm}  
  
  Given these definitions, we can usefully refine the notion of a ``small" violation:  Suppose one has an EFT with cutoff $\Lambda$ where all spacetime and gauge symmetries of the system have been identified.  Then we define a \emph{high-quality approximate global symmetry} to be one where the dimensionless coefficients (namely after appropriate powers of the cutoff have been extracted) of all symmetry-violating operators are \emph{exponentially small}.   The Swampland Global Symmetry Conjecture will bound just how high-quality any would-be global symmetry can possibly be, at least in the gauge-derived, type-(1) case.

  An important comment concerning `fine-tuned' or `type-(3)' global symmetries has to be made: We do not use the word `tuning' in 't~Hooft's sense \cite{tHooft:1979rat} since, of course, by its very definition the smallness of a coefficient is technically natural if its vanishing implies a global symmetry. Our main point is simply that an approximate symmetry might be present in the low-energy theory without a deep structural reason. Nevertheless, a relation to fine-tuning in the technical sense of 't~Hooft exists. Indeed, in the absence of an underlying gauge symmetry, there may and in general will be irrelevant operators violating the desired approximate symmetry. Thus, while having finitely many operator coefficients small at the irrelevant and marginal level does not need tuning in the low-energy EFT, the full theory will tend to correct those coefficients on the basis of its symmetry-violating UV structure. In this sense, a fine tuning will indeed be needed and the name `fine-tuned global symmetry' may be suitable in spite of the apparent clash with 't~Hooft's nomenclature.

  On a more general note, we should emphasize that the whole idea of conjecturing a quantum-gravity-derived minimal size of symmetry-violating effects goes against 't~Hooft's technical naturalness. The latter is a concept of QFT in non-dynamical spacetime and assumes that it is possible to arrange things so that the couplings of global-symmetry violating operators renormalize only multiplicatively.  We go beyond this by claiming that an unavoidable {\it additive} non-perturbative correction to the coefficients of such operators is always present.

 \subsection{Deriving the bound}
  
  Our focus will be on type-(1) or gauge-derived global symmetries. To explain our general logic, recall first the gauging of a $p$-form gauge theory by a $(p\!+\!1)$-form gauge theory (or equivalently the `Higgsing' of the latter by the former), see e.g.~\cite{Banks:2010zn}:
  \begin{equation}
   \frac{1}{g_p^2}|\diff A_p|^2+\frac{1}{g_{p+1}^2}|\diff A_{p+1}|^2\qquad \to \qquad
   \frac{1}{g_p^2}|\diff A_p+A_{p+1}|^2+\frac{1}{g_{p+1}^2}|\diff A_{p+1}|^2 \,.\label{gau}
  \end{equation}
  In the Higgsed version on the RHS, the charged $(p\!-\!1)$-branes of the $p$-form theory cease to exist as independent objects for lack of gauge invariance. They can only appear as boundaries of the $p$-branes charged under $A_{p+1}$:
  \begin{equation}\label{gaugeinvariance}
   S\supset \int_{B_p} A_{p+1}+\int_{\partial B_p}A_p \,.
  \end{equation}
  Only this combination is invariant under the gauge symmetry $\delta A_{p+1}= \diff\chi_p$, $\delta A_p=-\chi_p$ of the Higgsed model. (For simplicity, we ignore for now the option of introducing a relative integer factor between $\diff A_p$ and $A_{p+1}$ on the RHS of (\ref{gau}). 
  We comment on this below.)

\begin{figure}[t]
   \centering
   \includegraphics[width=0.6\textwidth]{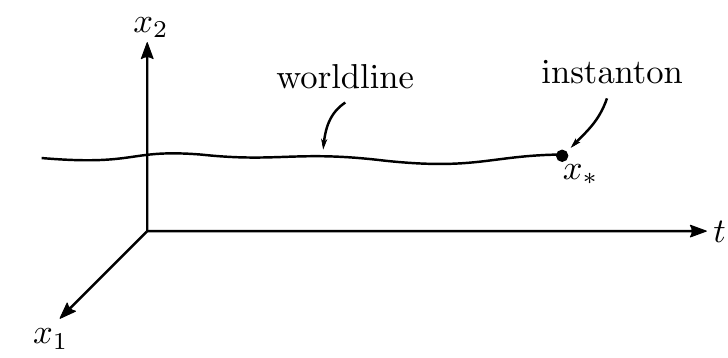}
   \caption{A worldline of a charged particle ending at an instanton at $x_*$. Integrating over instanton positions $x_*$ induces a global-charge-violating operator.}
   \label{fig:Instanton}
  \end{figure}

  Applied to our case of a 1-form Higgsed by a 0-form, this implies that instantons can only exist as the origin or endpoint of a worldline of a charged particle (see Fig.~\ref{fig:Instanton}):  
  \begin{equation}\label{eq:Coupling}
   S\supset \int_{B_1(x_*)}A_1\,\,+\,\,\phi(x_*) \,.
  \end{equation}
  Here $x_{*}$ is the location of an instanton and $B_1$ is the worldline of a light charged particle ending on it. It follows that the usual local (as far as the EFT is concerned) operator induced by the instanton sum also changes:
  \begin{equation}\label{gop}
   e^{-S_I+i\phi}\qquad\to\qquad \Phi\,e^{-S_I+i\phi} \,.
  \end{equation}
  Here, we for simplicity assumed that our light charged particle is a complex scalar $\Phi$, transforming as $\delta\Phi=\Phi e^{i\chi}$ for $\delta A=\diff\chi$ and $\delta\phi=-\chi$. 
  
  Now the instanton action $S_I$ is constrained by the WGC as $S_I \lesssim M_\Pl /f$ \cite{ArkaniHamed:2006dz}. Furthermore, as also argued in \cite{ArkaniHamed:2006dz}, the WGC in general also has a magnetic version, constraining the cutoff. Specifically in the axionic case and for parametrically small $f\ll M_\Pl$, both this magnetic WGC as well as a black-hole evaporation argument suggest that $\Lambda \lesssim \sqrt{f M_\Pl}$ \cite{Hebecker:2017uix}. This gives $S_I\lesssim M_\Pl^2/\Lambda^2$ and hence the desired bound of the coefficient $\alpha$ of the global-symmetry-violating operator in (\ref{gop}):
  \begin{equation}\label{eq:WeakGravityBreaking}
   \alpha \sim \exp\left(-S_I\right) \gtrsim \exp\left(- \frac{M_\Pl^2}{\Lambda^2}\right)\,.
  \end{equation}
  To be more precise, the factor $\exp(i\phi)$ makes the operator on the RHS of (\ref{gop}) gauge-invariant. But after gauge fixing to $\phi=0$, which is natural in the low-energy EFT, one is left with an instanton-suppressed global-$U(1)$-violating operator $\sim \alpha\,\Phi$, with the exponentially small coefficient $\alpha$ displayed above. Crucially, independently of any UV details, the WGC constrains the strength of the violation in terms of the cutoff of the 4d theory.
  
  At this point, an important comment has to be made: When gauging the 0-form theory with the $U(1)$, the axionic degree of freedom merges with the 1-form to produce a massive vector in the familiar manner. More generally, according to \eqref{gau} the $p$-form degrees of freedom merge with those of the $(p+1)$-form theory in the process of gauging.  A priori it could be that the WGC does not apply to the two theories independently in their respective forms when they couple in this way.  We here assume that it \emph{does}, similar to the use of the WGC in e.g.~\cite{Reece:2018zvv, Ibanez:2015fcv}. Strictly speaking this represents an additional assumption mildly generalizing the minimal WGC.
  
  We did not make a possible numerical coefficient in the exponent in (\ref{eq:WeakGravityBreaking}) manifest since, as long as we do not make precise what we mean by the cutoff $\Lambda$, such a coefficient can always be absorbed in the latter. However, as discussed in more detail in~\cite{Hebecker:2017uix}, the present cutoff is associated with the tension of strings (coupling to the 2-form dual to the axion) going to zero. It hence in general represents a much more fundamental breakdown of the EFT than just a finite set of new particle states appearing at some scale.  This situation also allows one in principle to make things more quantitative through replacing $\Lambda^2$ by the string tension $T_1$, such that $\exp(-M_\Pl^2/\Lambda^2)\,\,\to\,\,\exp(-c \, M_\Pl^2/T_1)$, where the ${\cal O}(1)$ coefficient $c$ could now in principle be determined.  This would require fixing the electric and magnetic versions of the WGC underlying our derivation exactly.\footnote{
   Precisely in the present case this is in fact non-trivial: On the one hand, it is not clear which object defines the WGC bound on the instanton side \cite{Heidenreich:2015nta, Hebecker:2016dsw, Hebecker:2018ofv} (options include the Giddings-Strominger wormhole \cite{Giddings:1987cg} or extremal instantons, which however involve a dilaton). On the other hand, the field-strength contribution to the tension of a charged string diverges in the IR, making also this side of the conjecture quantitatively more complicated \cite{Heidenreich:2015nta}. 
  } In the EFT, one might want to define the cutoff as $\Lambda = \min(m_A,\Lambda_A)$. Here, $m_A$ is the photon mass after Higgsing, $m_A = g \cdot f$ (see Sect.~\ref{sec:SimpleModels}), and $\Lambda_A$ is the cutoff set by the magnetic weak gravity conjecture, $\Lambda_A \lesssim M_\Pl/g$ \cite{ArkaniHamed:2006dz} for strong $U(1)$-coupling $g$ and therefore weak magnetic coupling $\widetilde{g} = 1/g$. (At weak coupling $g$ one finds $m_A \lesssim \Lambda_A = g M_\Pl$ by the WGC for axions.) One then finds $\Lambda \lesssim \sqrt{T_1}$ and therefore $\exp(-M_\Pl^2/\Lambda^2) \lesssim \exp(-M_\Pl^2/T_1)$. The strength of violation claimed in \eqref{eq:WeakGravityBreaking}, using this more general cutoff $\Lambda$, is therefore even weaker than the one we explicitly derived.

  Importantly, we expect that if an instanton-induced operator violates the global charge by one unit and is constrained as above, multi-instanton effects of instanton number $k$ will induce operators violating the global symmetry by $k$ units (here we are assuming for simplicity that the would-be conservation law is associated to a $U(1)$ and there is an additive quantum number) and be constrained to have coefficient above $\exp(-k \, M_\Pl^2/\Lambda^2)$. This structure of coefficients is consistent under renormalization group evolution of the EFT to lower scales. In addition if an operator with charge violation by $k$ units is induced by an instanton, loops will induce all other operators with the same degree of charge violation unless there are secretly further symmetries.\footnote{
   In the supersymmetric case this statement potentially requires some modification as there can be selection rules due to, e.g., holomorphy.  
  }
  If we now assume that the couplings of the theory not involved with the high-quality approximate global symmetry are not exponentially small, loop-suppression is non-exponential, and all operators violating the symmetry must appear with an overall structure of coefficients set by $\exp(-k \, M_\Pl^2/\Lambda^2)$ factors. 
  
  We have so far only focused on the exponential suppression of symmetry-violating operators. On top of it, there can be polynomial suppression by the cutoff.  This is for example the case, when symmetry-violating operators of a field $\Phi$, a $SU(N)$-singlet, are only loop-induced via coupling to a field $\psi$, a doublet under $SU(N)$ and which therefore couples to gauge instantons directly.  We expect all operators that are not forbidden by an additional (hidden) global symmetry to be loop-induced. This could for example exclude fermion mass terms, as these are usually protected by an additional global flavor symmetry respected by the 't~Hooft vertex\,\footnote{ 
   See however \cite{Caldi:1977rc,Carlitz:1977fn}, where instanton-induced operators generate fermion mass terms via loops at large gauge couplings.
  }. We comment on this further in Sect.~\ref{sec:Conclusion}.
  
  If we introduce an integer coefficient $n$ in the coupling of the two gauge sectors in (\ref{gau}), $\left|\diff \phi + n A_1\right|^2$, an unbroken $\mathbb{Z}_n\subset U(1)$ discrete gauge theory remains \cite{Krauss:1988zc,Alford:1989ch,Preskill:1990bm,Banks:2010zn}.  This gauge symmetry strongly constrains the allowed operators. Given that the lowest $U(1)$/$\mathbb{Z}_n$-charge can be normalized to $1$, we can bound the dimension of the smallest operator that breaks the global $U(1)$-symmetry to $d\leq n$\,: $\Phi^n e^{-S_I}$. If there are multiple fields with higher charges, already operators of smaller dimension can respect the $\mathbb{Z}_n$ gauge symmetry. 
  
  Finally, we note that a $U(1)$ global symmetry may be broken, e.g.~by a Higgs VEV, to a global discrete symmetry $\mathbb{Z}_n \subset U(1)$. In such cases, our bound \eqref{eq:WeakGravityBreaking} will of course apply to the latter.

 \section{Simple Models}\label{sec:SimpleModels}
 
 \subsection{A four dimensional example}
  
  We illustrate the above argument with a simple, explicit example that is UV-complete in 4d. By this we mean that our instantons are conventional gauge instantons, such that no point-like 0-dimensional objects need to be introduced. 
  
  As above our two ingredients are a $U(1)$ gauge theory with charged fermions $\psi$ on the one hand and an axion coupled to an $SU(N)$ gauge theory (and hence to instantons) on the other hand:
  \begin{equation}
   S_1=\int \diff^4x\left(-\frac{1}{e^2}F^2+\overline{\psi}i\slashed{D}\psi\right)\,,
   \quad 
   S_2=\int \diff^4x\left(-f^2(\partial\phi)^2-\frac{1}{g^2} \tr G^2 
   +\frac{\phi \,\tr G \widetilde{G}}{8\pi^2}\right)\,.
  \end{equation}
  We now gauge the axion, which so far only possesses the discrete gauged shift symmetry $\phi\to\phi+2\pi$, under the $U(1)$. Naively, one would simply replace $\partial_\mu \phi\to D_\mu \phi\equiv \partial_\mu \phi +A_\mu$. However, this is inconsistent due to the non-invariance of the last term in $S_2$ under gauge transformations $\delta\phi=\chi$. As explained in the general case, our gauging requires that charged worldlines end on instantons. In the case at hand, this can be realized by gauging the $U(1)$ charged fermions additionally under $SU(N)$.\footnote{
   We anyway expect that, as in the case of the WGC for multiple $U(1)$'s \cite{Cheung:2014vva}, there should be light states charged under both gauge groups.
  }
  
  The theory is then defined by
  \begin{equation}\label{eq:action}
   S=\int \diff^4x \left(-\frac{1}{e^2} F^2 - \frac{1}{g^2} \tr G^2 - f^2 (D\phi)^2 + \bar{\psi}_L i\slashed{D}_L \psi_L +\bar{\psi}_R i\slashed{D}_R\psi_R  +\frac{\phi \,\tr G \widetilde{G}}{8\pi^2}\right)\,.
  \end{equation}
  We have rewritten the Dirac fermion $\psi$ in terms of a l.h.~and a r.h.~spinor. Moreover, each of these has been promoted to an $SU(N)$ fundamental multiplet. For the theory to be free of a mixed $U(1)SU(N)^2$-anomaly, we impose the condition $q_L - q_R = 1$ on the $U(1)$-charges $q_L$ and $q_R$ of the left- and right-handed fermion.\footnote{
   Alternatively, we could have multiplied the $\phi G \widetilde{G}$ term by $(q_L - q_R)$.
  }
  
  We end up with a consistent theory\,\footnote{
   \,$U(1)$-gravity and $U(1)^3$ anomalies can be canceled by further St\"uckelberg terms (not involving the $SU(N)$) or by adding extra fermions, charged only under the $U(1)$ (for recent related work see \cite{Craig:2019zkf}). In both cases our main points below are not affected. Note that gravitational instantons contributing to the effective action via the $\phi R \widetilde{R}$-coupling  \cite{Delbourgo:1972xb,Eguchi:1976db} will parametrically give the same result as gauge instantons, see Sect.~\ref{sec:Wormholes}.
  }
  of fermions charged under $U(1) \times SU(N)$. Below the mass scale of the photon $A_\mu$, whose mass is induced by $f^2(D\phi)^2 \supset f^2 A^2$, the $U(1)$ appears only as a global symmetry.
  
  The $SU(N)$-instanton sum induces a 't~Hooft operator \cite{tHooft:1986ooh}, involving fermions and suppressed by $\exp(-S_I) = \exp(-8\pi^2/g^2)$, as part of the effective Lagrangian. In our case, it reads
  \begin{equation}
   \mathcal{O} = e^{-S_I} \, \bar{\psi}_L \psi_R \, e^{i \phi} + \text{h.c.}
  \end{equation}
  This operator is of course invariant under the $U(1)$ gauge symmetry thanks to the shift in the axion. However, once we gauge-fix the axion to $\phi = 0$ and remove it from the effective theory, the operator explicitly violates the global $U(1)$ which would have otherwise survived.
    
 \subsection{Comments on a possible relation to an effective axion}
  
  Until now we focused on fundamental axions with a coupling $\phi G \widetilde{G}$ in the microscopic Lagrangian. It is clearly interesting to ask whether our arguments, leading to the bound of \eqref{eq:WeakGravityBreaking}, can also be made in the case of an effective axion representing the phase of a complex scalar $H$. Indeed, let $H$ have an Abelian-Higgs-model potential, enforcing a non-zero VEV: $H=ve^{i\phi}\,$. Then the low-energy EFT only contains the effective axion $\phi$. The underlying global symmetry may be broken by operators $\alpha H + \bar{\alpha}\bar{H} \rightarrow \alpha v e^{i\phi}+\bar{\alpha} v e^{-i\phi}$, such that the full EFT partition function reads
  \begin{equation}
  \begin{aligned}
   \mathcal{Z} &=\int\mathcal{D}\phi\, \exp\left\{-S_0[\phi] + \int\left( \alpha v e^{i\phi} +  \bar{\alpha}ve^{-i\phi}\right)\right\}\\
   &= \int\mathcal{D}\phi\, \,e^{-S_0[\phi]} \,\, \sum_{n,\bar{n}=0}^\infty\,\,\frac{1}{n!\bar{n}!}\,\, \left(\int\alpha v e^{i\phi}\right)^{n}\left(\int\bar{\alpha} v e^{-i\phi}\right)^{\bar{n}}\,.
  \end{aligned}
  \end{equation}
  Here the second line can be interpreted as a sum over $n$-instanton/$\bar{n}$-anti-instanton sectors, which one would naturally expect to come with a fundamental axion.\footnote{
   Note that these terms do not originate in a $\phi G\widetilde{G}$ coupling to gauge instantons. We simply rearranged the perturbative terms in the path integral in a suggestive, instanton-like manner.
  }
  One then identifies $v\alpha$ with $e^{-S_I}$, such that the WGC for axions places a lower bound on the operator coefficient $\alpha$. This logic extends to the gauged case as follows: Include a $U(1)$ gauge theory which, according to the WGC, comes with a charged particle $\Phi$. To gauge $H$, we have to replace the `instanton-type' operator $\alpha H$ by $\alpha H\Phi^\dagger$. After integrating out the massive vector and axion, the low-energy EFT now contains the operator $\alpha v\Phi^\dagger$, corresponding to the destruction or creation of $\Phi$-particles (cf.~Fig.~\ref{fig:Instanton}), with a coefficient bounded from below by $e^{-M_\Pl/f}$.
  
  However, this Higgs-derived axion case may be different since our logic places only an exponentially small bound on operator coefficients which may be naturally ${\cal O}(1)$. In general, our Higgs field $H=v\,\exp(i\phi)$ is available for the construction of all kinds of operators in the high-scale theory. Thus, even in the fermionic case (where we previously had a symmetry reason for a light $U(1)$-charged particle) we must now allow for the operator $y\,H \bar{\psi}_L \psi_R$ to be present. Then no low-energy global $U(1)$ survives in the first place unless we can ensure that $y\ll 1$\,\footnote{
   A small value of $y$ may be technically natural in the 't~Hooft sense since the coupling $y$ can be forbidden by chiral symmetry. However, this further global chiral symmetry now comes in as an extra assumption. Our results limit how small $y$ may become. Fortunately, the condition that the charged particle appears in the low-energy effective theory is simply that $y$ is parametrically, not exponentially, small.
  }. In our present understanding, also more involved Higgs-based models of this type (with more fermions and other Higgs-charge) generically have the same feature: The survival of a global $U(1)$ before non-perturbative effects are included requires the choice of a small operator coefficient.\footnote{
   Exceptions of the Frogatt-Nielsen-type are possible at the price of having only a highly-charged field in the low-energy EFT (as discussed in \cite{Fichet:2019ugl} in the present context).
  }
  
  Let us pause to spell out the difference between the fundamental and effective axion cases at energies below the physical Higgs scale and before the axion is gauged by the 1-form $U(1)$ symmetry: Both cases are by definition built on a scalar with gauged discrete shift symmetry $\mathbb{Z}$. In both cases, shifts $\phi\to \phi+2\pi \epsilon$ with $\epsilon$ non-integer are {\it not} gauged and hence need {\it not} be respected by all terms in the Lagrangian. In the effective case, this simply means that one must suppress all higher-dimension operators violating such shifts by non-integer $\epsilon$. This requires additional tools, e.g.~tuning or extra symmetries. In the fundamental case, the standard form of the leading-order gauge theory action excludes non-derivative couplings of the axion. Indeed, our axion is viewed as a 0-form potential, the kinetic term is $|\diff \phi|^2$, and any further appearance of $\phi$ arises only in combination with charged objects. These are the instantons, allowing contributions with $\phi$ evaluated at their location, but only at the price of a factor $\exp(-S_I)$. A more fundamental reason for why this basic gauge theory structure can not be broken may be given as follows: We declare that a proper gauge theory must allow for both an electric and a magnetic formulation. Hence, an axion is {\it weakly-coupled fundamental} only if a dual 2-form description exists in which the instantons (now viewed a 0-dimensional defects enclosed by a quantized 3-form-field-strength integral) have action $S_I\gg 1$. In this dual formulation, local operators providing  non-derivative couplings of $\phi$ can not be written down. 

 \subsection{A simple five dimensional example}
  
  We now sketch a simple 5d toy model which contains some features of the string models quoted in Sect.~\ref{sec:EBranes}. 
  
\begin{figure}[t]
   \centering
   \includegraphics[width=0.7\textwidth]{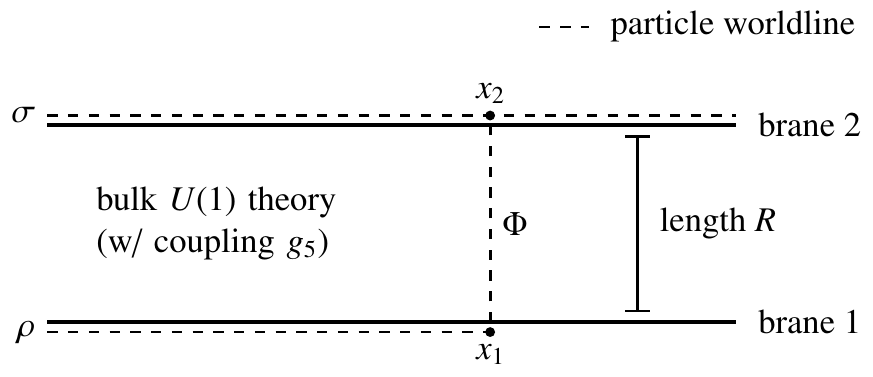}
   \caption{A 5d toy model with charged scalars confined to the two boundary-branes. In the presence of a $\sigma$-VEV, the bulk $U(1)$ gauge symmetry is broken but a global $U(1)$ survives.}
   \label{fig:5dmodel}
  \end{figure}

  Consider a $U(1)$ gauge theory on $\mathbb{R}^4\times S^1/\mathbb{Z}_2$. Let a charged scalar $\rho$ be localized on boundary 1 and, similarly, a scalar $\sigma$ on boundary 2. In addition, the WGC for the bulk $U(1)$ requires the existence of a charged bulk field $\Phi$ (see Fig.~\ref{fig:5dmodel}). Let $\Phi$ be a scalar for simplicity. We will return to the case where the WGC particle is a fermion at the end of this section. A VEV of $\sigma$\,, $\langle \sigma \rangle = v e^{i\theta} \neq 0$\,, gives a mass $m_A^2 = g_5^2 v^2 /R$ to the photon, leaving a global symmetry under which $\rho$ is charged at low energies.
  
  We assume that the couplings of $\Phi$ to both $\sigma$ and $\rho$ allowed by locality and gauge invariance are present. As a result, there are instanton-like processes in which charged $\rho$-particles disappear from their brane, see Fig.~\ref{fig:5dmodel}. Summing over all such `E0-brane instantons' induces a corresponding operator in the 4d effective action. It comes with a suppression factor $\exp(-S_\Phi)$, where $S_\Phi = m_\Phi \int_\text{E0} \mathrm{d}y \sqrt{-g} \propto m_\Phi R$ is the action of a Euclidean 0-brane stretched over the interval. Moreover, the coupling of $\Phi$ to the gauge field, $i \int_\text{E0} A\equiv i\phi$, gives a factor of $\exp(i \phi)$. All in all, this gives an effective gauge-invariant operator of the form 
  \begin{equation}
   \rho \, e^{-S_\Phi}e^{i \phi} \, \sigma = \rho \, e^{-S_\Phi}e^{i (\phi+\theta)}\, v\,.
  \end{equation}
  We may use the residual gauge symmetry to set $\phi+\theta=0$\,, which leaves us with a global-symmetry-breaking tadpole operator for $\rho$\,:
  \begin{equation}
   \mathcal{O}\,\sim\, v e^{-S_\Phi} \rho \,.
  \end{equation}
  From the 5d version of the WGC we have 
  \begin{equation}
   S_\Phi\,\sim\, m_\Phi R\lesssim \,g_5M_5^{3/2}R\,\lesssim\, M_5R\,,\label{m5e}
  \end{equation}
  where the last estimate uses the perturbativity requirement $g_5\ll 1/\sqrt{M_5}$. At first sight a reasonable cutoff of the 4d EFT is the compactification scale  $\Lambda = m_{\text{KK}}\sim1/R\ll M_5$. Then, using $R\sim M_4^2/M_5^3$, we find 
  \begin{equation}\label{eq:5dbound}
   S_\Phi\,\lesssim\, M_5R\,\sim\,\frac{M_4^2}{M_5^2}\,\ll\frac{M_4^2}{\Lambda^2}\,,
  \end{equation}
  which agrees with our general bound. We may try to go beyond this by raising the cutoff above $m_{\rm KK}$ and incorporating the (weakly coupled) tower of KK modes in the effective 4d description. A 4d EFT perspective may be maintained until the growing number of KK modes overcomes their weak coupling and one loses perturbative control. Unsurprisingly, this happens at the quantum gravity cutoff scale $\Lambda \sim M_5$ (see e.g.~\cite{Heidenreich:2017sim}). With this, the inequality on the r.h.~side of \eqref{eq:5dbound} is saturated.
  
  Finally, let us comment on the possible case that the WGC particle in the bulk is a fermion $\Psi$ of unit $U(1)$ charge. Since we only have a scalar on the brane, it is not possible to introduce a coupling between these two fields which would allow one unit of charge to be carried away from brane 1. However, a coupling $\sim(\rho^*)^2 \overline{\Psi}\Psi^c$ permits the removal of two units of charge and, correspondingly, an instanton-like process suppressed by two rather than one massive brane-to-brane propagators. More coupling options arise if one invokes the sub-lattice WGC and hence the presence of fermions with other charges. One may also postulate and apply a generalized version of the completeness conjecture using the charge lattice of $Spin(4,1)\times U(1)$. Here $Spin(4,1)$ acts on the bulk tangent bundle as prescribed by 5d relativity. Based on this, one can argue that both a charged scalar {\it and} a charged fermion must always be present at the 5d Planck scale, which is sufficient to derive \eqref{m5e}. The existence of charged bulk scalars and fermions also follows from the stronger conjecture of \cite{Palti:2020tsy} that supersymmetry should always be present at the energy scale set by the WGC. Either way, this logic makes our derivation of global-symmetry violation more general.

 \section{Direct quantum gravity and black hole arguments}\label{sec:QGExamples}
  
 \subsection{Gravitational instantons}\label{sec:Wormholes}
  
  Our arguments so far were indirect in that we used quantum gravity to support the WGC and the latter to argue for global symmetry violation. A more direct approach is the inclusion of gravitational instantons in the path integral. Most generally, we here mean contributions (ideally Euclidean solutions) with non-trivial 4d topology, as pioneered in \cite{Eguchi:1978xp,Eguchi:1978gw}. Among the many possible topological fluctuations (see e.g.~\cite{Hawking:1979zw,Hawking:1979hw}) the gluing of a K3 surface into $\mathbb{R}^4$ might be particularly interesting since its effect on fermions is quite analogous to the 't~Hooft vertex discussed earlier \cite{Hebecker:2019vyf}. This induces global symmetry violation, with the relevant operator suppressed by $\exp(-S_I)\sim \exp(-M_\Pl^2/\Lambda^2)$. The last expression follows simply from the facts that $M_\Pl^2$ multiplies the Einstein action and that the integral over instanton-sizes is dominated by the smallest objects allowed by the cutoff.
 
  A more general and maybe more intuitive way to violate global symmetries through topology change are Euclidean wormholes \cite{Giddings:1987cg, Lee:1988ge, Abbott:1989jw, Coleman:1989zu, Kallosh:1995hi}, which can be interpreted as a pair of gravitational instantons (each corresponding to the emission or absorption of a baby universe). This issue has in particular been recently revived in the context of the violation of global shift symmetries \cite{Montero:2015ofa, Hebecker:2015rya, Heidenreich:2015nta, Hebecker:2016dsw, Alonso:2017avz, Hebecker:2018ofv}.
 
  A parametric analysis of the Einstein-Hilbert action gives a suppression factor 
  \begin{equation}
   \exp(-S_I) \,\,\sim\,\,  \exp(-M_\Pl^2 L^2) \label{whs}
  \end{equation}
  for wormhole-induced global symmetry violation. Here $L$ is the typical wormhole radius. The smaller the wormhole,  the smaller is the action and the weaker the suppression in \eqref{whs}. Requiring the wormhole to be controlled in the EFT implies $L\gtrsim1/\Lambda$, which saturates the bound \eqref{eq:WeakGravityBreaking}.
  
  Let us now look more specifically at a gauge-derived global symmetry and its possible violation by a wormhole with topology $\mathbb{S}^3\times I$, where $I\subset \mathbb{R}$ is an interval. The worldline of our $U(1)$-charged particle passing through the wormhole fills out $I$ and is a point in the $\mathbb{S}^3$. As shown in App.~\ref{sec:AppendixA}, consistency of the gauged axion theory then requires that the $\mathbb{S}^3$ also carries an $H_3$ flux equal to the particle's charge. Here $H_3$ is the 3-form field strength dual to the axionic 1-form field strength $\diff\phi$. Let us make the assumption (to be justified momentarily) that the wormhole radius is $L \sim 1/\sqrt{f M_\Pl}$, as in the pure Giddings-Strominger case, without gauging or charged particle. Here, $f$ is the axion decay constant. On the length scale $L$, the photon is effectively massless, $m_A=ef\ll 1/L$, as long as the gauge coupling $e$ remains perturbative. But since the 2-form $B_2$ has the same mass, $m_B=m_A$, one may then treat it as massless too. Hence, the Giddings-Strominger analysis \cite{Giddings:1987cg} still holds, justifying our assumption $L\sim 1/\sqrt{fM_\Pl}$ above. A slightly more detailed argument is given in App.~\ref{sec:AppendixA}. Treating the regime of large $e$ is beyond the scope of this work.
  
  Note that we are interested in situations where the globally charged particle is visible to the low-energy observer, i.e. $m\ll 1/L$. Then the above picture of a classical, localized particle and hence of a charge-carrying worldline passing through the wormhole (with its topological implications) is insufficient. However, the opposite limit where the particle's charge is completely smeared over the transverse $\mathbb{S}^3$ leads to the same conclusion: The $H_3$ flux still supports the wormhole as in the pure Giddings-Strominger case. More details are again given in App.~\ref{sec:AppendixA}. We leave a full quantum treatment and the study of the intermediate regime of such a fundamentally new wormhole solution for future work.
  
  Finally, we note that there are open fundamental questions related to Euclidean wormholes. In particular, there may be problems with the definition of the Euclidean path integral for gravity in general as well as deep conceptual issues with the summation over baby universe states in particular (see \cite{Hebecker:2018ofv} for a review and \cite{Marolf:2020xie, McNamara:2020uza, Gesteau:2020wrk} for recent developments). We also note, that unlike gauge theories where cluster decomposition can be used to argue that a sum over non-trivial gauge topologies must be included in the path integral, there is no analogous definitive argument that gravitational instantons, or more general configurations of non-trivial topology must be part of the gravitational path integral \cite{Witten:1985xe}. Nevertheless, we view the agreement between the old wormhole/gravitational-instanton logic and the WGC-based derivation noteworthy.
  
 \subsection{String constructions and Euclidean branes}\label{sec:EBranes}
  
  If quantum gravity is defined by string theory, one may appeal to the precise (though not general) arguments against exact global symmetries of  \cite{Banks:1988yz}. The situation is even better in AdS space: Inconsistency of global symmetries can be proven using properties of the dual CFT \cite{Harlow:2018tng}. Clearly, it is non-trivial to map this to realistic string (or AdS/CFT derived) models with non-perturbative effects, broken supersymmetry and positive vacuum energy. Nevertheless, explicit constructions of global symmetries (e.g.~\cite{Ibanez:2001nd, Antoniadis:2002cs, Ibanez:2006da}) support what was said in Sects.~\ref{sec:MainArgument} and \ref{sec:SimpleModels}:
  
  For global symmetries arising from gauge symmetries on branes in string compactifications\,\footnote{
   Concretely, $U(1)$-anomalies of D-branes gauge theories are canceled by a 4d version of the original Green-Schwarz mechanism. The gauge boson acquires a St\"uckelberg mass and the symmetry survives as a perturbatively exact global symmetry. It may then only be violated by non-perturbative effects.
  }, it has been established that Euclidean D-brane instantons induce symmetry-violating operators \cite{Blumenhagen:2006xt, Ibanez:2006da, Florea:2006si, Blumenhagen:2009qh}\,\footnote{
   See \cite{Martucci:2015dxa, Martucci:2015oaa} for constructions in M- and F-theory.
  }. These operators are governed by a coefficient 
  \begin{equation}
   \exp(-S_\text{E$p$}) \sim \exp\left(- T_p \text{Vol}(\Sigma_{p+1})\right) \sim \exp\left(-\frac{1}{g_s} \left(\frac{R}{l_s}\right)^{p+1}\right) \sim \exp(-M_\Pl/f)\,.
  \end{equation}
  Here $\Sigma_{p+1}$ is the cycle wrapped by the E$p$-brane (with tension $T_p$), $R$ is a typical compactification scale and we have suppressed all numerical coefficients. The last relation involves estimating the decay constant $f$ of a $C_{p+1}$-axion coupling to the E$p$-brane.
  
  Taking the cutoff $\Lambda$ to be the KK-scale, $m_\text{KK}\sim 1/R$, we arrive at
  \begin{equation}\label{eq:branemodelscale}
   \exp\left(-S_\text{E$p$}\right) \sim \exp\left(-\frac{M_\Pl^2}{\Lambda^2} \, g_s \left(\frac{l_s}{R}\right)^{7-p}\right) \,.
  \end{equation}
  We see that in the perturbative regime, $g_s \ll 1$ and $l_s/R\ll1$ (and $p\leq 5$ for purely internal Euclidean branes), the suppression factor is generically much smaller and the violation therefore much stronger than \eqref{eq:WeakGravityBreaking}. This is because we typically find $\Lambda \ll \sqrt{f M_\Pl}$, that is, the magnetic version of the axionic WGC is not saturated by $m_\text{KK}$. Of course, this is not surprising: While $m_\text{KK}$ is the obvious and maybe most reasonable cutoff to apply to the 4d EFT the cutoff may also be raised by including KK modes in the 4d description (see the discussion around \eqref{eq:5dbound}). Therefore it is expected that $m_{\rm KK}$ as a cutoff does not saturate the fundamental WGC. We also recall that, in the stringy context, there is a well known smooth transition between brane instantons and the pure gauge-theory instantons we discussed in our earlier toy model.
  
  We close by emphasizing that our generic bound does not become uninteresting just because stringy models have well-understood brane instantons. Indeed, one could try to perfection a global symmetry by considering very special geometries and brane arrangements, hoping to achieve an arbitrarily high quality of the symmetry from the 4d EFT perspective. If our 4d derivation can be established, such attempts would be known a priori to be futile. 
   
 \subsection{Black hole effects in a thermal bath}\label{sec:ThermalBath}
  
  While black holes are maybe the origin of our conviction that quantum gravity violates global symmetries, it is not obvious how to relate their effect to the desired operator coefficients. A recent suggestion made in \cite{Fichet:2019ugl} is based on a `local rate bound'. The latter says that in a thermal bath with $T\lesssim \Lambda$ the violation rates of a global symmetry should obey
  \begin{equation}\label{eq:LocalRateBound}
  \Gamma_{\text{BH}}\lesssim \Gamma_{\text{EFT}}\,.
  \end{equation}
  Here, $\Gamma_{\text{BH}}$ and $\Gamma_{\text{EFT}}$ are the charge violation rates induced by thermal black hole fluctuations (dominated by black holes with $R_{\text{BH}}\sim\Lambda$) and by local operators explicitly included in the EFT, respectively. 
 
  A possible conjecture (different from~\cite{Fichet:2019ugl} -- see below) is then that {\it any EFT coupled to gravity with cutoff $\Lambda$ and possessing an approximate global symmetry should satisfy \eqref{eq:LocalRateBound}.} While the motivation of \eqref{eq:LocalRateBound} remains mysterious to us,\footnote{
   Arguments in favor of this bound  which assume particle-like objects in the energy domain above the cutoff have recently been discussed in v2 of \cite{Fichet:2019ugl}.
  } such a conjecture would be intriguing by its simplicity and attractive implications: One easily derives from it a bound of the type
  \begin{equation}\label{eq:FSBound}
   \alpha \gtrsim \exp(-M_\Pl^2/\Lambda^2)
  \end{equation}
  for the coefficient $\alpha$ of the operator which dominates \eqref{eq:LocalRateBound} \cite{Fichet:2019ugl}. This is in fact immediately obvious if one considers the thermal black hole abundance $\sim \exp(-M_{\rm BH}/T)\sim \exp(-M_{\rm BH}/\Lambda)$ together with the smallest allowed black hole mass $M_{\rm BH}\sim M_\Pl^2 R\sim M_\Pl^2/\Lambda$. Moreover, one may write the action for a black hole propagating for a time $\tau$ as
  \begin{equation}
   M_{\rm BH}\,\tau \,\,\sim\,\, \int \diff^3x\int_0^\tau    \diff t\, M_\Pl^2\, \sqrt{-g}{\cal R} \,\,\sim\,\, R^3\,\tau\, M_\Pl^2\,\frac{1}{R^2}\,,
  \end{equation}
  and use this as an estimate of the black hole mass. Then it becomes apparent that the above derivation of \eqref{eq:FSBound} fits perfectly in the scheme underlying all bounds discussed in this paper:
  
  In the end, in all cases the number in the exponent is just the factor $M_\Pl^2$ of the Einstein-Hilbert action, with the $1/\Lambda^2$ supplied on dimensional grounds. One way or the other, one appears to rely on a topology fluctuation of size $1/\Lambda$. In the WGC-version, this is hidden in the WGC bound on instantons, but it is secretly still present in that wormholes saturate that bound. 
  
  Unfortunately, things are not that simple and the conjecture proposed in~\cite{Fichet:2019ugl} is in fact much weaker. In our interpretation, it says that in any EFT with cutoff $\Lambda_1$ it should be possible to raise the cutoff to a scale $\Lambda_2\geq \Lambda_1$ such that \eqref{eq:LocalRateBound} holds. One reason for this is the existence of clockwork-style $N$-field gauge theories in which all operators up to mass dimension $\sim e^N$ respect a certain global $U(1)$. Given a species-bound-motivated cutoff $\Lambda\sim M_\Pl/\sqrt{N}$, it is clear that the symmetry-violation rates scale as $\exp[-\exp(M_\Pl^2/\Lambda^2)\ln(\Lambda)]$, such that the local rate bound can be violated. One way out is to accept that one may have to raise the cutoff, such that the full lattice~\cite{Heidenreich:2015nta} of charged states comes into play. Then symmetry-violating operators of lower mass-dimension become accessible and the rate bound is respected. Another option would be to conjecture that the required clockwork-style \cite{Choi:2015fiu, Kaplan:2015fuy} models will not be found in the landscape. Crucially, a conjecture for which one may need to raise the cutoff has limited use for the low-energy observer. It may also be possible to break it by adding a sector with light strings at some scale between $\Lambda_1$ and $\Lambda_2$, which formally stops one from raising the 4d cutoff.

 \section{Synthesis of results}\label{lh}
  
 \subsection{Comments and a possible loophole}\label{lh1}
  
  Of the three types of approximate global symmetries we discussed (gauge-derived, accidental and fine-tuned), our focus was on the first case. Here, we provided a general argument bounding the exponential suppression of symmetry-violating operators (cf.~\eqref{eq:WeakGravityBreaking} of Sect.~\ref{sec:MainArgument}) However, we did not address the important issue of non-exponential prefactors. Indeed, a (simplified) generic form of a symmetry-violating operator is
  \begin{equation}
   \mathcal{O} = \, C \,  e^{-c\, M_\Pl^2/\Lambda^2} \, M_\Pl^4 \left(\frac{\Lambda}{M_\Pl}\right)^k \left(\frac{\Phi}{M_\Pl}\right)^d \,,
  \end{equation}
  with real numbers $C$ and $c$ as well as an integer $k$. While we exclude $C \ll 1$ by demanding that the operator is not fine-tuned, the suppression by hierarchies in scales can be strong. Our ignorance of non-exponential coefficients derives not only from instanton prefactors, but also from the loop effects to which we appeal when claiming that operators with different dimensions and field-content can be loop-generated on the basis of a single instanton-induced operator. 
  
  Furthermore, there is a loophole (related to potentially large numerical coefficients in the exponent) whose resolution might come with interesting new insights into the nature of instantons or the Weak Gravity Conjecture\,: Let us assume that in our underlying $U(1)$ gauge theory (which obeys the completeness hypothesis) all fields with charges $q=1,\ldots,k-1$ are heavy, $m_i \sim \Lambda$, and only the field $\Phi_k$ with charge $k$ is light. Then any EFT operator violating the global symmetry must do so by $k$ units. It hence derives from a $k$-instanton effect and is correspondingly suppressed: $\Phi_k \exp(-k \, S_I)$. For the observer in the EFT, the global symmetry is much more precise than expected since he cannot know that $\Phi_k$ is highly charged in the underlying gauge theory. Possibly, this is resolved once one includes gravitational instantons: Since all fields couple to gravity, there will be operators suppressed by the gravitational instanton action $\exp(- S_I)$ without any further parameters. Alternatively, it may be impossible to make all the lower-charge fields parametrically heavier than $\Phi_k$. This option is interesting since it also works towards inhibiting the method of breaking of the WGC in the low-energy EFT by Higgsing \cite{Hebecker:2015rya,Saraswat:2016eaz}.
  
  We note that such a model is subject to strong consistency constraints if the spectrum is fermionic as in our prime example of Sect.~\ref{sec:SimpleModels}: Hierarchies in fermion masses affect the available fermion spectrum that has to cancel $U(1)$-gravitational and $U(1)^3$ anomalies \cite{Craig:2019zkf}. This might further constrain the cutoff of the anomaly-free theory, or, turning the argument around, the fermions that we are allowed to make heavy without the low-energy theory becoming inconsistent.
  
  But maybe the most straightforward way of dealing with this loophole is by insisting that we are in the setting of non-tuned, `type-(1)' global symmetries: In other words, we have to insist on a symmetry reason for the lightness of $\Phi_k$. This requires that $\Phi_k$ transforms in a non-trivial representation of some group $G$. If $G$ is identical to our gauged $U(1)$, underlying the global $U(1)$ we are discussing, then $\Phi_1$ is made light by the same argument as $\Phi_k$. This is what happens for the chiral fermions of Sect.~\ref{sec:SimpleModels}. By contrast, if $G$ is some further gauged or (gauge-derived) global symmetry, then we expect by completeness that a charged field $\Phi_1'$ exists. This field should have unit charge under our basic $U(1)$ and transform under $G$ just like $\Phi_k$. Then we expect that the symmetry argument keeping $\Phi_k$ light also applies to $\Phi_1'$. As a result, the low-energy observer would see symmetry breaking effects associated with $\Phi_1'$ and suppressed only by $\exp(-S_I)$, closing the potential loophole.
  
 \subsection{\texorpdfstring{Towards formulating a general\\ Swampland Global Symmetry Conjecture}{Towards formulating a general Swampland Global Symmetry Conjecture}}\label{lhg}
  
  Our argument could be the starting point for a derivation of a Swampland Global Symmetry Conjecture, but there are a number of caveats. To see this, we need to recall our classification of global symmetries in three categories: gauge-derived, accidental, and fine-tuned. There are now different possible conjectures to be made:
  
  First, we could be modest and satisfied with the fact that our constraint applies only to global symmetries of the gauge-derived type.
  
  Second, we could recall that our logic (allowing also for loop effects) in fact suggests that {\it all} operators are affected by our constraint: Any operator violating the $U(1)$ charge by $n$ units comes with a prefactor generically not smaller than $\exp(-nS_I) \sim \exp(- n M_\Pl^2/\Lambda^2)$. Beating this by a landscape-type tuning is clearly impossible in a finite landscape, so we can conjecture as follows: An EFT where, for any $n$, all operators violating a global $U(1)$ by $n$ units have coefficient below $\exp(- n M_\Pl^2/\Lambda^2)$ is in the swampland. This is very general but also very weak since it can be satisfied by an operator of very high mass dimension.
  
  Third, we could take an operator-focused approach and declare that accidental global symmetries simply {\it do not count} as approximate global symmetries. This is not as unnatural as it seems since, in a model with accidental global symmetry, low-dimension operators violating the symmetry are forbidden by gauge invariance. Thus, if one defines EFTs with an approximate global symmetry as those where the {\it coefficients of symmetry-violating operators are unnaturally small}, accidental symmetries are excluded: in these, the missing operators simply do not exist. With this, one may now hope that a stronger conjecture holds for the remaining gauge-derived and fine-tuned symmetries: There exists some $\Lambda_0$ such that no EFT with $\Lambda<\Lambda_0$ has any operator violating a global symmetry by $n$ units with coefficient below $\exp(- n M_\Pl^2/\Lambda^2)$.  Clearly, establishing this requires knowledge about the tuning-power of the landscape and its growth with $M_\Pl/\Lambda$.\footnote{
   To be more precise, one would have to allow for a prefactor $f_n(M_{\rm P}/\Lambda)$ and constrain its form. Moreover, it is clearly conceivable that the truth lies somewhere in between demanding that at least one symmetry-violating operator or all such operators satisfy our bound.
  }
  
  Fourth, we could maintain the above form of the conjecture for gauge-derived and tuned symmetries without excluding accidental symmetries from consideration. In this case, we would have to postulate a separate bound for symmetry violation in accidental global symmetries, as suggested in~\cite{Fichet:2019ugl} using a simple model and the WGC. Unfortunately, the suggested bound on the maximal mass-dimension up to which all operators can be forbidden may be evaded by clockwork-type EFT constructions. One would then need to hope that the latter are in the swampland.
  
  We also note that a universal statement about global symmetries of all types has been suggested \cite{Fichet:2019ugl} on the basis of a `local rate bound' in a thermal plasma (see our Sect.~\ref{sec:ThermalBath}). To avoid  the above problem of `clockworked accidental-symmetries', the authors formulate their conjecture in a fairly weak, UV-sensitive way (the cutoff may need to be raised to see that a certain EFT satisfies the bound). As we just explained, both in the gauge-derived case and possibly more generally, we would like to claim a stronger bound, purely in the low-energy EFT. However, we are also faced with a field-content-related loophole, as we now recall:
  
  One can imagine scenarios in which a particle of charge $k$ under the originally gauged $U(1)$ is light while all the lower-charge particles are heavy. As explained in Sect.~\ref{lh1}, this light particle can only disappear at the price of a $k$-instanton effect, suppressed by $\exp(-k M_\Pl^2/\Lambda^2)$. Thus, while we gave arguments in Sect.~\ref{lh1} for why such a peculiar mass arrangement might be impossible to realize, at least for $k\gg 1$, our bounds remain conjectures (even taking the WGC for granted).

 \section{Conclusion}\label{sec:Conclusion}
  
  Let us start by summarizing our fundamental point as presented in Sect.~\ref{sec:MainArgument}: A gauge-derived global $U(1)$ symmetry can arise if a gauged $U(1)$ is Higgsed by an axion.  This only requires that some charged particles survive below the St\"uckelberg mass scale. Now, since the axion unavoidably couples to instantons and the latter, equally unavoidably, violate global $U(1)$ charge, we can quantify the global-symmetry violation in the low-energy EFT.  More precisely, the electric and magnetic form of the WGC for axions constrain the instanton action in terms of the cutoff, leading to an upper bound $\exp(-S_I)\gtrsim \exp(- M_\Pl^2/\Lambda^2)$ for the relevant dimensionless operator coefficients.  Moreover, the cutoff $\Lambda$ is related to the tension of the string associated with our axion theory.
  
  Our argument could be the nucleus for a derivation of a Swampland Global Symmetry Conjecture, and we demonstrated a number of supporting examples in Sects.~\ref{sec:SimpleModels} and \ref{sec:QGExamples}, but there are a number of caveats and potential loopholes, as discussed in Sect.~\ref{lh}.    We also presented in Sect.~\ref{sec:MainArgument} a classification of global symmetries in three categories: gauge-derived, accidental, and fine-tuned.  It could logically be the case that our bound applies only to gauge-derived global symmetries without the accidental or fine-tuned mechanisms also being operative.   However our logic suggests a number of other possibilities for a general Swampland Global Symmetry Conjecture, as we briefly discussed in Sect.~\ref{lhg}.  
  
  There are further interesting questions left open by our analysis. For example, we presented a generalization of the Giddings-Strominger wormhole solution to the case of a gauge-derived $U(1)$ global symmetry where a charged particle passes through the wormhole.  This leads to a violation of the global symmetry of size in accord with our bound. However the (possible) role of wormholes and other topologically non-trivial configurations in a quantum theory of gravity is still far from settled, and merits much more study.
  
  In summary, while we believe to have made progress in developing a Swampland Global Symmetry Conjecture, there are clearly many interesting open issues that remain to be resolved.

 \vspace{1cm}
  
 \phantomsection
 \subsection*{Acknowledgments} 
  We would like to thank Joerg Jaeckel and Timo Weigand for useful comments. This work is supported by the Deutsche Forschungsgemeinschaft (DFG, German Research Foundation) under Germany's Excellence Strategy EXC 2181/1 - 390900948 (the Heidelberg STRUCTURES Excellence Cluster) and  the  Gra\-du\-ier\-ten\-kol\-leg `Particle  physics  beyond  the  Standard  Model' (GRK  1940).
  
 \vspace{1cm}

 \appendix
  
 \section{The wormhole solution with charged particles} \label{sec:AppendixA}
  
  Consider a wormhole of topology $\mathbb{S}^3 \times I$  and a particle of unit charge under a gauge-derived global $U(1)$ passing through it. Let the particle sit at the north pole of $\mathbb{S}^3$. The wormhole dynamics is best understood using the magnetic dual description of the axion theory. The magnetic coupling is $f$ and the (Euclidean) action in our gauged case is
  \begin{equation}
   S=S_{\text{EH}}+\int\left(\frac{1}{f^2}H_3\wedge\star H_3+e^2\mathcal{F}_2\wedge\star\mathcal{F}_2\right)\,.
   \label{aact}
  \end{equation}
  Here $\mathcal{F}_2=\widetilde{F}_2+B_2$ is a gauge-invariant magnetically dual $U(1)$ field strength (with magnetic coupling $e^{-1}$), cf.~\eqref{gau} for $p=1$. Away from electrically charged particles, we have $\diff \mathcal{F}_2 = H_3$.
  
  Consider, in complete generality, a smooth patch of $\mathbb{R}^4$ and a worldline of a unit-charge particle passing through it. This charge is measured by
  \begin{equation}
   \int_{\mathbb{S}_\epsilon^2}\mathcal{F}_2 = \int_{\mathbb{S}_\epsilon^2}\widetilde{F}_2=1\,,
  \end{equation}
  where $\mathbb{S}_\epsilon^2$ is a sphere of infinitesimal radius $\epsilon$ threaded by the worldline. Here the first equality follows since by assumption $B_2$ is smooth in our patch.
  
  Now, consider a 3-sphere of our wormhole with an infinitesimal ball $B_\epsilon$, centered on the north pole, cut out: $\mathbb{S}^3\backslash B_\epsilon$. We find
  \begin{equation}
   \int_{\mathbb{S}^3\backslash B_\epsilon} H_3 = -\int_{\mathbb{S}_\epsilon^2} \mathcal{F}_2 = -1 \,,
  \end{equation}
  where the sign signals the different orientation of the boundary of $\mathbb{S}^3\backslash B_\epsilon$ relative to $\mathbb{S}_\epsilon^2$.
  
  The interpretation of this is as follows: If a $U(1)$ gauge theory is Higgsed by an axion, then a $U(1)$-charged particle traveling through a wormhole must be accompanied by an appropriate $H_3$ flux. That is, the field $\mathcal{F}_2$ induced by the particle is compensated for by flux on the rest of the sphere. This analysis appears to support the persistence of the Giddings-Strominger solution \cite{Giddings:1987cg} in the Higgsed case. 
  
  However, the above arguments were purely topological. To understand the dynamical solution, let us write the relevant part of the action on $\mathbb{S}^3\backslash B_\epsilon$ in the form
  \begin{equation}\label{eq:EnergyDensities}
   S~\supset~ \int\, \frac{1}{f^2}\Big[ |H_3|^2 + e^2 f^2 |\mathcal{F}_2|^2\Big]~=~ \int \,\frac{1}{f^2} \Big[ |\diff B_2|^2 +  m_B^2 |B_2|^2\Big]\,.
  \end{equation}
  Here in the last expression we have chosen the gauge $\widetilde{F}_2 = 0$. This is always possible on $\mathbb{S}^3\backslash B_\epsilon$ since there is no charged particle. Let us view the geometry as fixed and of typical size $L$ and try to understand the $B_2$ solution. Since we work classically, the overall prefactor $1/f$ of the action may be ignored. Then our problem has two terms whose ratio is governed by the mass parameter $m_B=m_A=ef$. We expect the term suppressed by $m_B$ to be irrelevant in a geometry of size $L$ if $m_B\ll 1/L$. But this last relation holds in our regime of interest where the electric coupling is weak, $e \lesssim 1$, and $L\sim 1/\sqrt{fM_P}$. Thus, the $H_3$ flux spreads out approximately homogeneously on the sphere, as in the pure Giddings-Strominger case. The perturbation of the field profile by the charged particle is negligible.\footnote{
   We recall, for the convenience of the reader, that the wormhole radius then follows by demanding that the flux energy density, $|H_3|^2/f^2 \sim 1/(f^2 L^6)$, equals the gravitational energy density, $M_\Pl^2 R \sim M_\Pl^2/L^2$. Solving this for $L$ gives $L\sim 1/\sqrt{fM_P}$, as quoted earlier.
  }
  
  In the opposite regime, $m_B \gg 1/L$, the first term in \eqref{eq:EnergyDensities} would dominate such that the field $B_2$ should strongly localize around the particle at the cost of large field-gradients. A quantitative discussion of this regime goes beyond the scope of this work.
  
  Finally, as already pointed out in the main text, the arguments above rely on the presence of a localized charged-particle worldline in the wormhole. But in our regime of interest the particle's Compton wavelength is larger than the wormhole radius, $m\ll 1/L$. So a better model may be that of an electric charge distribution $j_3$ smeared homogeneously over the transverse $\mathbb{S}^3$. While an action principle for the magnetic fields $\widetilde{F}_2=\diff\tilde{A}_1$ in the presence of a smooth electric current is notoriously hard to formulate, the equations of motion are easy to write down:
  \begin{equation}
   \diff\widetilde{F}_2=j_3\,\,,\qquad \diff \star \widetilde{F}_2=0\,.
  \end{equation}
  Here we assume $j_3$ to be proportional to the volume form on $\mathbb{S}^3$. It is immediately clear that $B_2=-\widetilde{F}_2$,  ${\cal F}_2=0$, extremizes the action (\ref{aact}) together with a harmonic 3-form flux $H_3$ that is homogeneously distributed as in the solution by Giddings and Strominger:\footnote{
   Note that both $B_2$ and $\widetilde{F}_2$ are gauge dependent and can only be defined locally with $B_2 = -\widetilde{F}_2$ on every patch. We refrain from introducing patches in \eqref{eq:ChargeFluxBalance}.
  }
  \begin{equation}\label{eq:ChargeFluxBalance}
   \int_{\mathbb{S}^3}H_3=\int_{\mathbb{S}^3}\diff B_2= -\int_{\mathbb{S}^3}\diff\widetilde{F}_2= -\int_{\mathbb{S}^3}j_3=-1\,.
  \end{equation}
  
  In summary, a Euclidean wormhole solution supported by 3-form flux persists even if $B_2$ is Higgsed and the charged particle passing the wormhole is light. We highlight that this is an extension of the Giddings-Strominger solution by a charged particle which couples to the Giddings-Strominger axion as specified implicitly in \eqref{aact}. This solution therefore differs from the extensions found in \cite{Abbott:1989jw, Coleman:1989zu}, where the axion was coupled to only an additional real field to form a complex scalar field.

  \bibliographystyle{JHEP}
  \bibliography{references}
  
\end{document}